# Imaging the high-frequency charging dynamics of a single impurity in a semiconductor on the atomic scale

Maialen Ortego Larrazabal[1,2,*], Jiasen Niu[2,3,7,*], Stephen R. McMillan[4,5], Paul M. Koenraad[6], Michael E. Flatté[4,6], Milan P. Allan[2,3,7, §], Ingmar Swart[1, §]

1 Debye Institute of Nanomaterials Science, Utrecht University, PO Box 80000, 3508 TA Utrecht, The Netherlands.
2 Leiden Institute of Physics, Leiden University, Niels Bohrweg 2, 2333 CA Leiden, The Netherlands.
3 Fakultät für Physik, Ludwig-Maximilians-Universität, Schellingstrasse 4, München 80799, Germany.
4 Department of Physics and Astronomy, University of Iowa, Iowa City, Iowa 52242, USA.
5 Donostia International Physics Center (DIPC), 20018 Donostia–San Sebastián, Spain.
6 Department of Applied Physics and Science Education, Eindhoven University of Technology, 5612 AZ Eindhoven, The Netherlands.
7 Munich Center for Quantum Science and Technology (MCQST), München, Germany.

* These authors contributed equally
§ Email: i.swart@uu.nl and milan.allan@lmu.de

**As electronic devices approach the atomic limit, the charge dynamics of individual dopant atoms increasingly constrain performance, stability, and coherence. In scanning tunnelling microscopy (STM), donor ionization is typically interpreted as a static threshold process arising from tip-induced band bending. Here we show that the ionization of individual sulfur donors in InAs is intrinsically dynamic and governed by the local electric field. Using MHz-frequency STM noise spectroscopy with atomic-scale spatial mapping, we resolve pronounced random telegraph noise that is invisible in time-averaged tunnelling spectra. A bias-dependent model quantitatively links the noise spectra to microscopic ionization and neutralization processes of the donor states, enabling direct extraction of nanosecond charge-state lifetimes. The switching rate is strongly bias dependent, demonstrating that the electric field continuously drives charge-state transitions. Unexpectedly, we show that the degenerately doped bulk leads to a sharp bias-dependent onset of donor ionization as the donor level crosses the Fermi level, giving rise to a characteristic shoulder in the noise power spectrum that is captured by our model. These results establish donor ionization as a non-equilibrium dynamical process with nontrivial contribution by the bulk electrons, and identify impurity switching as a universal nanoscale charge-noise mechanism relevant to quantum devices.**

As semiconductor technology advances, both by increasing material purity and by shrinking device dimensions, the influence of individual dopants becomes increasingly important[1]. Understanding how the local atomic structure, chemical composition and charge dynamics of a semiconductor affect its electronic properties is therefore crucial, not only for classical electronics but also for emerging quantum technologies, where single dopants can act as qubits or strongly affect coherence times[2,3]. The dynamics of dopant charging along with spin correlations drives spin blockade readout schemes for qubits[4,5] as well as some all-electrical room-temperature quantum sensors[6].

Individual impurities can affect the device behavior, e.g., by introducing states inside the bandgap. Dynamic effects, often attributed to the trapping and detrapping of carriers at localized impurity sites, manifest as random telegraph signals, also known as two-level fluctuations (TLFs), in the device current or conductance. A detailed understanding of the origin and dynamics of TLFs associated with single atom impurities in semiconductors is therefore essential to improve the performance of classical semiconductors, as well as semiconductor-based quantum technologies.

Random telegraph noise (RTN) signals have been extensively studied in mesoscopic transport (magnetic tunnel junctions[7,8], semiconducting quantum dots[9–11], MOSFETs[12,13], Josephson junctions[14]). Studies that correlate TLF dynamics with the local electronic landscape and atomic configurations, on scales available from scanning tunneling microscopy (STM) [15], would clarify the mechanisms of these effects and help connect them to mechanisms for room temperature electrically-detected magnetic resonance[16] and spin bottleneck[16,17] in modern semiconductor devices in small magnetic fields. STM has been extensively used to map out the spatial extent of impurity-induced charge fluctuations and to uncover how local electrostatics influence the dynamics of dopants at the nanoscale [18–23].

However, because of the limited bandwidth of conventional STM electronics, only processes with relatively slow dynamics can be studied using this technique. Moreover, low-frequency spurious noises, such as $1/f$ noise and mechanical vibrations, further complicate the detection. To overcome the bandwidth limitations of standard STM, we implemented high-frequency electronics[24,25]. This provides us with a detection bandwidth in the MHz range, making it possible to directly resolve rapid two-level charge fluctuations associated with individual dopants, while retaining atomic spatial resolution.

In this study, we employ MHz-frequency scanning tunnelling microscopy (STM) to probe high-frequency two-level fluctuations (TLFs) of individual dopant atoms below the surface of clean InAs. Spatial variations in the measured tunnelling-current noise are directly linked to tip-induced donor ionization, establishing a quantitative connection between random telegraph noise (RTN) and the underlying charge-transfer processes. To interpret these fluctuations, we develop a bias-dependent physical model that captures the competing ionization and neutralization processes of the charge-states of the donors under the local electric field of the STM tip. By combining noise spectroscopy with this model, we extract intrinsic lifetimes of the neutral and charged states on the nanosecond timescale, in excellent agreement with theoretical expectations for donor-electron dynamics. These rapid stochastic processes, inaccessible in conventional STM current-time traces, are thus revealed and quantitatively characterized, providing a direct window into high-frequency single-dopant charge kinetics. To arrive to these findings, we use InAs, which is a narrow bandgap III-V semiconductor in which sulfur atoms on arsenic sites are expected to introduce shallow donor states in the bandgap[26]. At cryogenic temperatures, dopant atoms can be ionized by the tip of the STM beyond a threshold voltage applied between the tip and the sample. Due to the poor screening of charge in these materials, the electric field induced by the voltage applied to the STM tip penetrates tens of nanometers into the InAs, bending the energy bands near the surface[27,28]. Depending on voltage and distance to the dopant, the tip induced band-bending (TIBB) can be strong enough to ionize impurities[29]. Figure 1 illustrates this process for a donor impurity. When the tip is far from the donor, the impurity remains neutral (Fig. 1a-b), as the thermal energy at 4.2 K is not sufficient to ionize it (Fig. c-d).

As the tip approaches the donor site (Fig. 1e-f), the local band bending can push the donor level above the Fermi energy, causing an electron of the dopant to move into the conduction band, thereby ionizing the atom. This ionization typically occurs within a small region, just a few nanometers laterally and in depth. The size of this ionization zone, that appears as a disc shaped area in topographic images and in STS images as a ring shape [19,21], depends on parameters such as the tip shape, tip-sample distance, bias voltage, and the local dielectric constant[28].

Ionization of the dopant atom locally reduces the effective tunneling barrier height, leading to an increase in the tunneling current. This appears as a small step in the I-V curve (top panel of Fig. 1g), indicating the voltage at which the donor becomes ionized for a specific tip-to-donor distance (green curves in Fig. 1g). The differential conductance (dI/dV) spectrum of such an I-V as a function of bias is presented at the bottom panel of Fig. 1g. A small peak in the dI/dV spectrum marks the onset of ionization, signifying the transition from a neutral to a charged state.

The dopant charging induced by TIBB is fully reversible: when the electric field is reduced, or when the tip is moved laterally away from the site, the impurity returns to its neutral state. In addition to this controlled neutralization, brief neutral periods can also occur spontaneously due to the stochastic arrival of electrons from the bulk conduction band. This dynamic interplay between ionization and relaxation forms the core mechanism behind the random telegraph noise (RTN) observed near donor impurities[21,30,31]. In traditional STM experiments, RTN manifests as two (or more) distinct levels in the current-time signal. However, given the limited bandwidth of conventional STM electronics, only dynamic processes with frequencies up to a few kHz can be detected this way. As we show below, accessing the power-spectral density of the tunnelling current makes it possible to probe much faster processes.

**Model for Random Telegraph Noise:**

We now discuss how charge state fluctuations affect the power spectral density of the tunnel current. Consider a dopant atom switching between the neutral and positive charge state, which can be viewed as a two-level system with stochastic fluctuations. The average lifetimes of the two states ($\tau_0$ for the neutral and $\tau_+$ for the charged) vary with experimental parameters such as applied bias and tip-sample distance.

The measurement frequency ($f$) - dependent power spectral density of the current signal is given by [32,33]

$$S_{RTN}(f) = \frac{S_0 \tau_{eff}}{1+4\pi^2 f^2 \tau_{eff}^2}, \tag{1}$$

With $S_0$ the integrated noise power, and $\tau_{eff}$ the effective lifetime of the switching events. These parameters depend on the average lifetimes $\tau_0$ and $\tau_+$ in the following way:

$$S_0 = 4(\Delta I)^2 \frac{\tau_{eff}}{\tau_0+\tau_+} \text{ and } \frac{1}{\tau_{eff}} = \frac{1}{\tau_0} + \frac{1}{\tau_+}, \tag{2}$$

Where $\Delta I$ is the amplitude of the two-level current fluctuations. The lifetime of the neutral state, $\tau_0$, is given by the ionization rate $\Gamma_+$, which we model as a field-driven tunnel escape through a tilted Coulomb potential using the Ammosov-Delone-Krainov (ADK)[34,35] tunnelling formalism (details in Methods II). Specifically,

$$\tau_0 = \tau_0^0 \exp\left(\frac{\beta}{V - V_{0/+}}\right)\left[1 + e^{\eta(\mu'-V)}\right], \tag{3}$$

where $\tau_0^0$ is the bias-independent attempt time (intrinsic lifetime), $\beta$ is a proportionality constant set by the tunnelling barrier parameters and $\eta$ and $\mu'$ are proportionality constants related to the binding energy of the donor and the Fermi energy of the bulk. $V_{0/+}$ is the bias voltage at which switching begins to occur.

On the other hand, the lifetime of the ionized state, $\tau_+$, is set by the neutralization (refilling) rate $\Gamma_0$, which we model as a thermally-assisted process of electrons tunnelling in from the bulk of the InAs. That takes the form:

$$\tau_+ = \tau_+^0 \exp\left(\alpha(V - V_{0/+})\right), \tag{4}$$

With $\tau_+^0$ as the intrinsic lifetime and $\alpha$ a proportionality constant that contains the temperature-dependence of the refilling process. For more information about the model see Methods II.

In effect, the two lifetimes follow distinct bias-voltage dependencies: $\tau_0$ decreases rapidly for increasing bias above $V_{0/+}$ (because the tunnelling barrier is lowered and the escape rate increases), while $\tau_+$ increases with bias (because the effective defect activation energy rises and the thermal refilling slows). Consequently, as the bias increases, the average lifetimes of both states shorten ($\tau_0$ becomes shorter, $\tau_+$ becomes longer, but the overall fluctuation cycle speeds up), so that the switching events become more frequent until $\tau_0$ becomes vanishingly small. The current-noise level is therefore governed by the density of such switching events per unit time, whereas the time-averaged (DC) current is proportional to the fraction of time spent in each state, which in turn depends directly on $\tau_0$ and $\tau_+$.

Figure 2 illustrates the the time evolution of the two-level fluctuations in the current due to the ionization process at different bias voltages (Fig. 2a), and the calculated spectral density of the current noise as a function of bias for a given frequency (Fig. 2b). In these figure $I_0$ represents the current level corresponding to the neutral state (Fig. 2c), and $I_+$ represents the current level at the charged state (Fig. 2d). Further details of the calculations are provided in the Supplementary Information.

The evolution of the current fluctuations due to the ionization process as a function of bias voltage goes through several stages. At low bias (i), the donor impurity remains neutral, and the current shows no signs of telegraph fluctuations; as a result, the current noise does not contain any contribution from impurity switching. As the bias increases (ii) band bending aligns the donor level with the conduction band, allowing the impurity to become ionized (Fig. 2c). This occurs when an electron is emitted into the conduction band. Because the n-type InAs bulk is degenerate[26], the ionization rate is strongly suppressed until the donor level is lifted above the Fermi energy. This sharp onset is captured by the Fermi–Dirac occupation factor in Eq. (3), which introduces a step-like increase in the ionization rate and gives rise to the characteristic shoulder observed in the noise power spectral density, as shown in in Fig. 2b(ii). Once ionization becomes allowed, electrons can also tunnel back from the conduction band, neutralizing the donor. The impurity therefore begins to switch randomly between charged and neutral states (Fig. 2d). At this stage, the characteristic lifetimes of the two states, $\tau_0$ and $\tau_+$, are more similar, leading to a further increase in the current noise. As the bias continues to increase (iii), the neutral state's average lifetime $\tau_0$ decreases significantly, crossing the point in which $\tau_+ \sim \tau_0$, resulting in faster switching and a higher telegraph noise amplitude. Finally, at very high bias values (iv), the electric field is so strong that the impurity can no longer be neutralized, $\tau_+ \gg \tau_0$, resulting in a decrease in observed current noise.

**Equivalence between RTN measurements and spectroscopic maps:**

To measure the current noise associated with the charging dynamics of a single dopant atom, we use a low-temperature STM equipped with a dedicated home-built amplifier operating at MHz frequency (see Supplementary Information Methods III). All noise measurements are performed at a constant junction resistance, where the noise power is expected to scale linearly with bias voltage (S = 2eV/R) due to the shot noise contribution[36–38]. In the presence of two-level fluctuations, however, the bias-dependent noise signal is expected to show the characteristic feature indicated in Fig. 2b on top of a linear background (which can be subtracted – see Supplementary Information Methods III).

The topographic images in Figs. 3a,b,d, together with the lattice sketch in Fig. 3c, identify individual subsurface dopants and the surface atomic lattice of InAs(111)A. At a bias of 200 mV, the spatial map of differential conductance (Fig. 3f) reveals a distinct 'goggle-shaped' feature, which we attribute to two closely spaced donor impurities (D1 and D2), in agreement with the line cut shown in Fig. 3e (see Supplementary Information Fig. S1 for the evolution of the differential conductance ionization maps as function of bias voltage). The corresponding noise map at the same bias (Fig. 3i) shows enhanced current-noise intensity that co-localizes with the conductance feature, indicating that the noise enhancement arises from impurity ionization. At a fixed position the dI/dV spectrum plotted as a function of bias voltage (Fig. 3g) shows a small peak that marks the onset of ionization (neutral to charged state). The random-telegraph-noise (RTN) spectrum recorded at the same location (Fig. 3h) captures the resulting power spectral density of the switching events between the neutral and ionized states. With this, we establish a clear relation between impurity ionization driven by tip-induced band bending and the observed RTN arising from switching between the ionized and neutral states.

The analysis of the S(V) curves shows the characteristic behaviour expected of a random-telegraph-noise (RTN) signal (Fig. 4a,b). From this we extract the average lifetimes of the neutral and charged states (Fig. 4c,d). In the S:InAs system, we find at the ionization bias voltage, $\tau_0 = 0.01\ \mu s$ and $\tau_+ = 0.01\ \mu s$. The fit yields the bias-independent intrinsic lifetimes $\tau_0^0$ and $\tau_+^0$, which lie in the nanosecond regime. These values correspond to switching frequencies in the gigahertz range, consistent with processes involving rapid electron impingement[39]. In our interpretation, $\tau_0^0$ (charged to neutral) is governed by tip-induced ionization, whereas $\tau_+^0$ (neutral to charged) is likely controlled by electrons of the bulk InAs bands refilling the impurity. This agrees well with the MHz to GHz frequencies attributed to electron impinging processes. This interpretation aligns well with the GHz switching rates characteristic of electron-impingement phenomena [39].

In agreement with the limited bandwidth of the feedback electronics, we do not observe two-level fluctuations (TLFs) in low-frequency STM current time traces.

Our ability to measure the dynamics of two-level systems in the sub-$\mu s$ range, undetectable in standard time-domain current traces, grants access to a previously unexplored regime of impurity dynamics. In contrast to the slow, thermally activated transitions typically observed in traditional transport or noise spectroscopy, the fluctuations we observe likely originate from the fast changes in the tunneling barrier height due to the electronic environment caused by the donor dopant, which is affected by the local electric field of the STM tip. The strong spatial correlation between enhanced current noise and ionization rings observed in STS maps further supports this electrostatic coupling and points to field-assisted charge switching mechanism.

**Conclusions:**

In conclusion, we demonstrate that local noise spectroscopy using MHz-frequency STM measurements can resolve fast two-level fluctuations associated with the charging of individual impurities in a semiconductor. These fluctuations are both bias- and position-dependent, and they exhibit a strong spatial correlation with tip-induced ionization features seen in STS. Importantly, our results reveal that even when the tunneling current appears stable, it can host hidden charge dynamics with nanosecond timescale. These fast impurity charging fluctuations are invisible in conventional current measurements but are directly revealed in the current noise. By combining the measurements with a bias-dependent physical model that captures the competing ionization and neutralization pathways of the donors, we are able to quantitatively extract nanosecond charge-state lifetimes and link the noise directly to microscopic charge-transfer processes. Combined with atomic-scale spatial resolution, this approach not only helps our understanding of atomic-scale charge transport but also opens new avenues for exploring the quantum behavior of dopant-induced fluctuations in semiconductors. This could provide a powerful route for instance for identifying and controlling individual charge fluctuations that limit the stability and coherence of quantum-dot and donor-based qubits[3,40], as well as clarify the mechanisms of electrically-detected magnetic resonance and spin bottleneck in low magnetic fields. This would pave the way for the development of next-generation electronic devices with improved performance and reliability.

**Data availability:**

All data used to generate the figures in the main text and the supplementary information is available upon request to the authors.

**Code availability:**

The code used for this project is available upon request to the authors.

**Acknowledgements:**

The authors acknowledge Daniel Vanmaekelbergh and Thomas Gozlinski for useful discussions. This work was supported by the European Research Council (ERC StG SpinMelt, ERC CoG PairNoise and ERC CoG Fractal (865570)). I.S. acknowledges the research program "Materials for the Quantum Age" (QuMat) for financial support. This program (Registration Number 024.005.006) is part of the Gravitation program financed by the Dutch Ministry of Education, Culture and Science (OCW). Theoretical development of the model (MEF) was supported by the US Department of Energy, Office of Basic Energy Sciences, under Award DE-SC0016379. S.R.M. acknowledges the Agencia Estatal de Investigación MCIN/AEI/10.13039/501100011033 through Proyecto de Generación de Conocimiento PID2023-146694NB-I00 (GRAFIQ), the European Union NextGenerationEU/PRTR-C17.I1, as well as by the IKUR Strategy under the collaboration agreement between Ikerbasque Foundation and DIPC on behalf of the Department of Education of the Basque Government.

**Figures:**

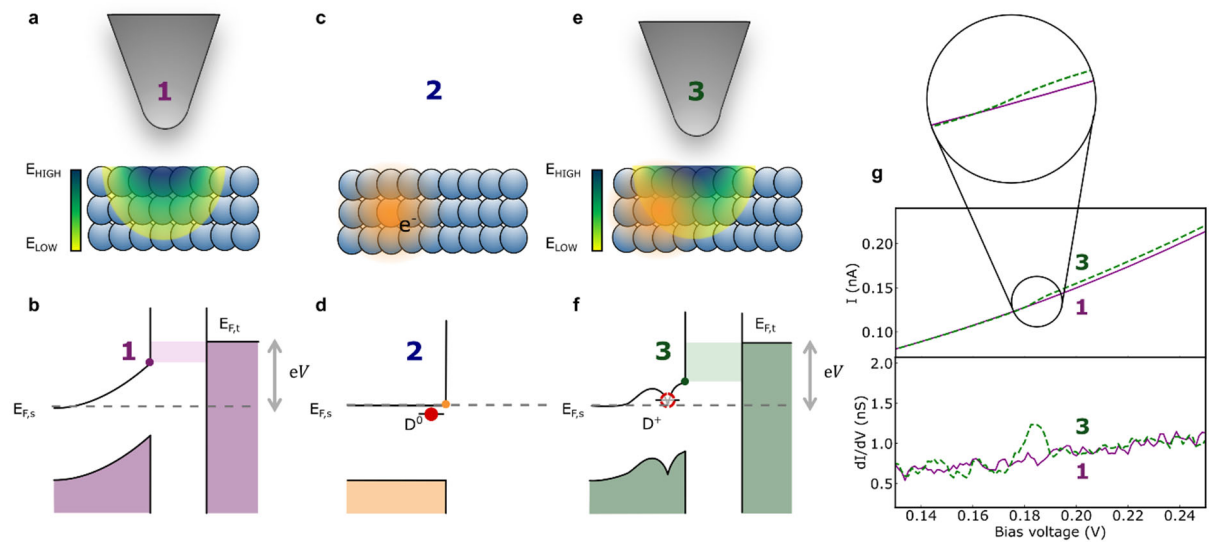

**Figure 1 | Tip-induced band bending and impurity ionization.** Schematic representation of the ionization process due to the tip-induced band-bending **a** – **f** and corresponding spectroscopic characteristics **g**. Band diagrams of the tip-sample junction (**b, d, f**) indicating the tip and sample Fermi energies ($E_{F,t}$ and $E_{F,s}$, respectively) and the bias voltage (eV). The electric field generated by the tip lifts the bands following the TIBB effect (case 1, sketch in a, band diagram in **b**). The neutral donor state coming from an impurity close to the surface lies below the $E_{F,s}$ (sketch in b, band diagram in **d**). As the tip approaches a donor, at a sufficient positive bias voltage, the donor state aligns with the conduction band and the electron escapes, ionizing the donor (case 3, sketch in e, band diagram in **f**). This process has an effect in the I-V curve (top panel in **g**), where the current is larger for the same biases when the donor is ionized, due to the increment in the available tunnelling states. In the dI/dV this effect corresponds to a peak in the signal (bottom panel of **g**).

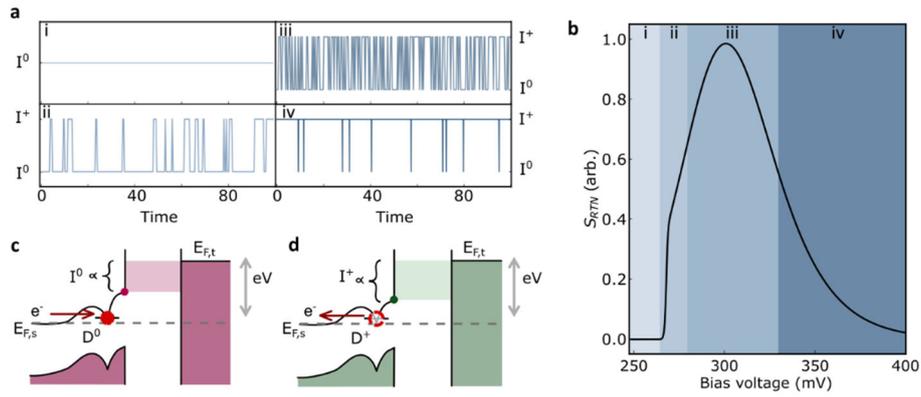

**Figure 2 | RTN model. a**, Four regimes (i–iv) encountered as the bias voltage increases. The current $I^0$ refers to the neutral state, and $I^+$ to the charged state. **b**, The resulting random telegraph noise (RTN) amplitude as a function of bias voltage, computed using the model in Equation (1). Further explanations of the processes at each regime are explained in the text. The shaded zones in **b** correspond to the regimes illustrated in **a**, showing how the switching dynamics influence the observed RTN signal. **c, d** Illustrative cartoons of the dynamic transitions driving the system into the **c** neutral state or **d** charged states.

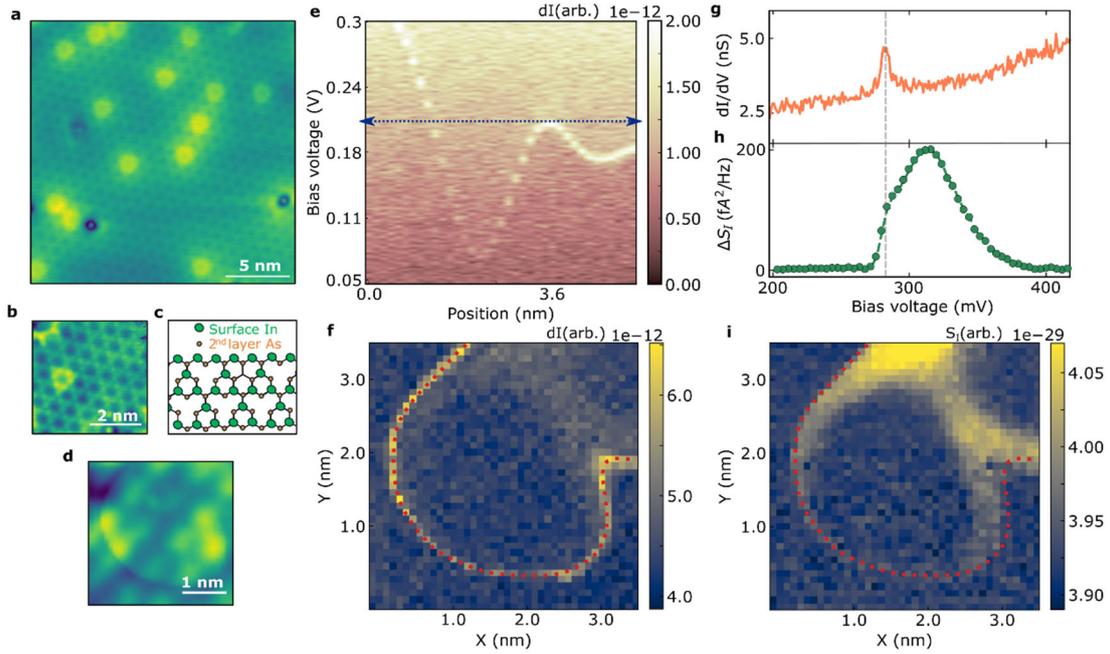

**Figure 3 | Relation between impurity ionization and RTN noise signal. a** Topography of the InAs(111)A surface over a 20 × 20 nm area. Bright regions are attributed to dopants in layers near the surface; darker regions correspond to depletion rings around In adatoms on the surface. **b** Zoomed-in topography of a 5 × 5 nm area, revealing a hexagonal pattern of the surface lattice formed by the In atoms. **c** Schematic sketch of the surface lattice and the first As layer. **d** Topography near a dopant, showing the characteristic disk-like feature associated with ionization. **e** dI/dV line-cut as a function of position across a region containing two dopants; color indicates dI intensity. The arrow marks the energy for which the spectroscopic map in **f** is taken. **f** Spectroscopic map (bias = 200 mV, $R_J = 250$ MΩ) over a 3.5 × 3.5 nm area; color indicates dI intensity. **g** Single-position spectrum (setpoint: V = 450 mV, I = 1.8 nA) as a function of bias voltage. **h** Corresponding RTN (random telegraph noise) spectrum measured at the same position as **g** ($R_J$ = 250 MΩ). **i** Noise map acquired under the same conditions as in **f**; the red dotted line traces the trend of the ionization ring seen in the spectroscopy map.

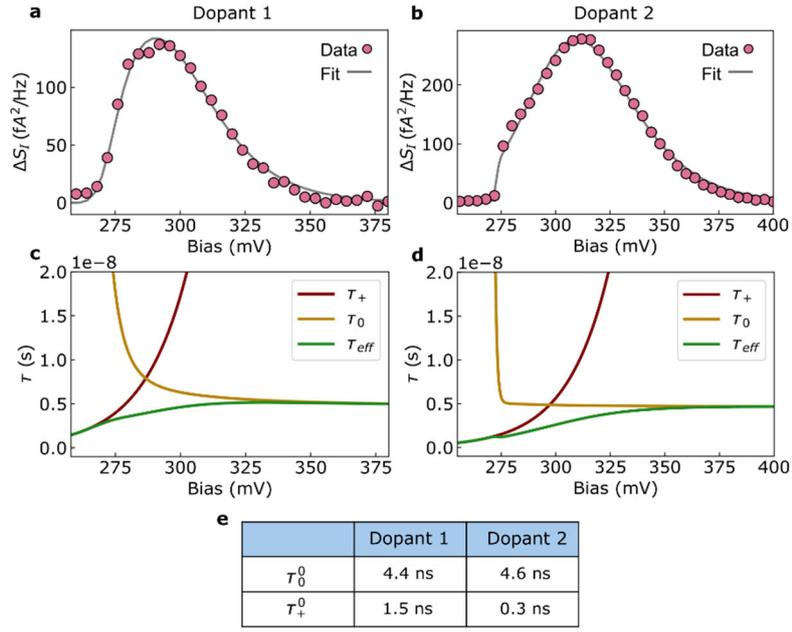

**Figure 4 | Measured RTN signal as function of bias voltage and fits. a,** RTN measured for a constant $R_j = 100$ MΩ (pink markers) noise measurement and the corresponding fit using the RTN equation (1) (grey solid line) for dopant 1. **b,** same measurement for dopant 2, taken at $R_j = 180$ MΩ. **c** and **d** are the corresponding lifetimes as function of bias voltage obtained from the fits in **a** and **b**, respectively. **e,** values of the intrinsic lifetimes at the ionization bias ($\tau_0^0$ and $\tau_+^0$) for two closely spaced impurities extracted from the fits. The 95% confidence values (see Methods V for details of the fit confidence bands) are $\tau_0^0 \in [3.64, 5.18]$ ns and $\tau_+^0 \in [1.30, 1.76]$ ns for dopant 1, and $\tau_0^0 \in [4.52, 4.82]$ ns, and $\tau_+^0 \in [0.25, 0.40]$ ns for dopant 2.